\documentclass[aps,prl,reprint,superscriptaddress]{revtex4-1}   

\usepackage{amsmath,amsfonts,amssymb}
\usepackage{wrapfig}
\usepackage{graphicx}
\usepackage{bbm}
\usepackage{float}
\usepackage{color}
\usepackage{braket}
\usepackage{bm}
\usepackage[unicode=true,colorlinks=true,citecolor=blue,urlcolor=blue]{hyperref}

\begin{document}

\title{
Spin and reoccupation noise in a single quantum dot\\ beyond the fluctuation-dissipation theorem}

\author{Julia Wiegand}
\affiliation{Institut f\"ur Festk\"orperphysik, Leibniz Universit\"at Hannover, Appelstra\ss{}e 2, D-30167 Hannover, Germany}
\author{Dmitry S. Smirnov}
\affiliation{Ioffe Institute, Polytechnicheskaya 26, 194021 St. Petersburg, Russia}
\author{Jens H\"ubner}
\email{jhuebner@nano.uni-hannover.de}
\affiliation{Institut f\"ur Festk\"orperphysik, Leibniz Universit\"at Hannover, Appelstra\ss{}e 2, D-30167 Hannover, Germany}
\author{Mikhail M. Glazov}
\affiliation{Ioffe Institute, Polytechnicheskaya 26, 194021 St. Petersburg, Russia}
\author{Michael Oestreich}
\email{oest@nano.uni-hannover.de}
\affiliation{Institut f\"ur Festk\"orperphysik, Leibniz Universit\"at Hannover, Appelstra\ss{}e 2, D-30167 Hannover, Germany}

\date{\today}


\begin{abstract}
We report on the nonequilibrium spin noise of a single InGaAs quantum dot charged by a single hole under strong driving by a linearly polarized probe light field. The spectral dependency of the spin noise power evidences a homogeneous broadening and negligible charge fluctuations in the environment of the unbiased quantum dot. Full analysis of the spin noise spectra beyond the fluctuation-dissipation theorem yields the heavy-hole spin dynamics as well as the trion spin dynamics. Moreover, the experiment reveals an additional much weaker noise contribution in the Kerr rotation noise spectra. This additional noise contribution has a maximum at the quantum dot resonance and shows a significantly longer correlation time. Magnetic-field-dependent measurements in combination with theoretical modeling prove that this additional noise contribution unveils a charge reoccupation noise which is intrinsic in naturally charged quantum dots.
\end{abstract}


\maketitle



The efficient optical spin manipulation of individual two-level systems opens fascinating perspectives for spin-photon interfaces, quantum cryptography, and quantum information processing \cite{Imamoglu1999,Loredo2017}. One of the particularly promising solid state candidates for such spin-photon quantum devices is a single spin localized in a semiconductor quantum dot (QD). Such a system not only provides very long coherence times and a large spin-photon coupling strength, but also provides controlled tuning of the emission wavelength and of the fine-structure splitting.
Indeed efficient spin manipulation~\cite{Giesz2016,DeGreve2011,Greilich2006}, spin detection~\cite{PhysRevB.78.085307,Arnold2015a}, and generation of highly indistinguishable photon pairs \cite{Somaschi2016,Dousse2010,Guerreiro2014} were already demonstrated in optimized quantum dot microcavity structures.

The spin dynamics of semiconductor quantum optical systems can be optimally studied by spin noise spectroscopy (SNS)~\cite{aleksandrov81} which avoids nonresonant optical excitation. The quantum optical technique has been transferred to semiconductor physics during the past decade~\cite{Oestreich_noise,Berski-fast-SNS,crooker2010,crooker2012,hackmann2015,Dahbashi2012}, revealing not only charge and nuclear spin dynamics~\cite{Berski2015b,OpticalField}, but also higher-order spin correlators~\cite{SNS_Universality,4order_exp}.
SNS is mostly used as a weakly interacting, nondestructive measurement technique. In this case the  fluctuation-dissipation theorem relates the spin noise spectrum to the magnetic spin susceptibility of the system~\cite{ll5_eng,NoiseGlazov} and thus strongly limits the informative abilities of the equilibrium SNS. Spin noise (SN) measurements beyond the fluctuation-dissipation theorem, i.e., under conditions out of thermal equilibrium, require a special theoretical analysis, but they can offer more information about the coupling and correlation between spin coherences, the response to resonant driving fields, and the charging dynamics~\cite{noise-trions,Mollow-noise,Nonresonant_nonequilibrium,Horn2011}. Lately, the first SN of a single QD was reported~\cite{Dahbashi2014a}, where the QD resonance was \emph{inhomogeneously} broadened due to slow charge fluctuations of residual impurities in the surrounding of the QD~\footnote{Since the timescale of charge fluctuations is much longer than the spin relaxation time, this has the same effect as inhomogeneous broadening in an ensemble of QDs.}.
In this Rapid Communication we employ SNS in view of spin-photon interfaces and investigate the spin dynamics of a coherent superposition of a single hole and the corresponding trion state of a \emph{homogeneously} broadened strongly driven QD. The observed dynamics in the coupled QD microcavity system allow us to understand physical limits and challenges of optically driven charged QDs as solid-state qubits--one of these challenges being the intrinsic reoccupation noise.

\begin{figure}
\includegraphics[width=\columnwidth]{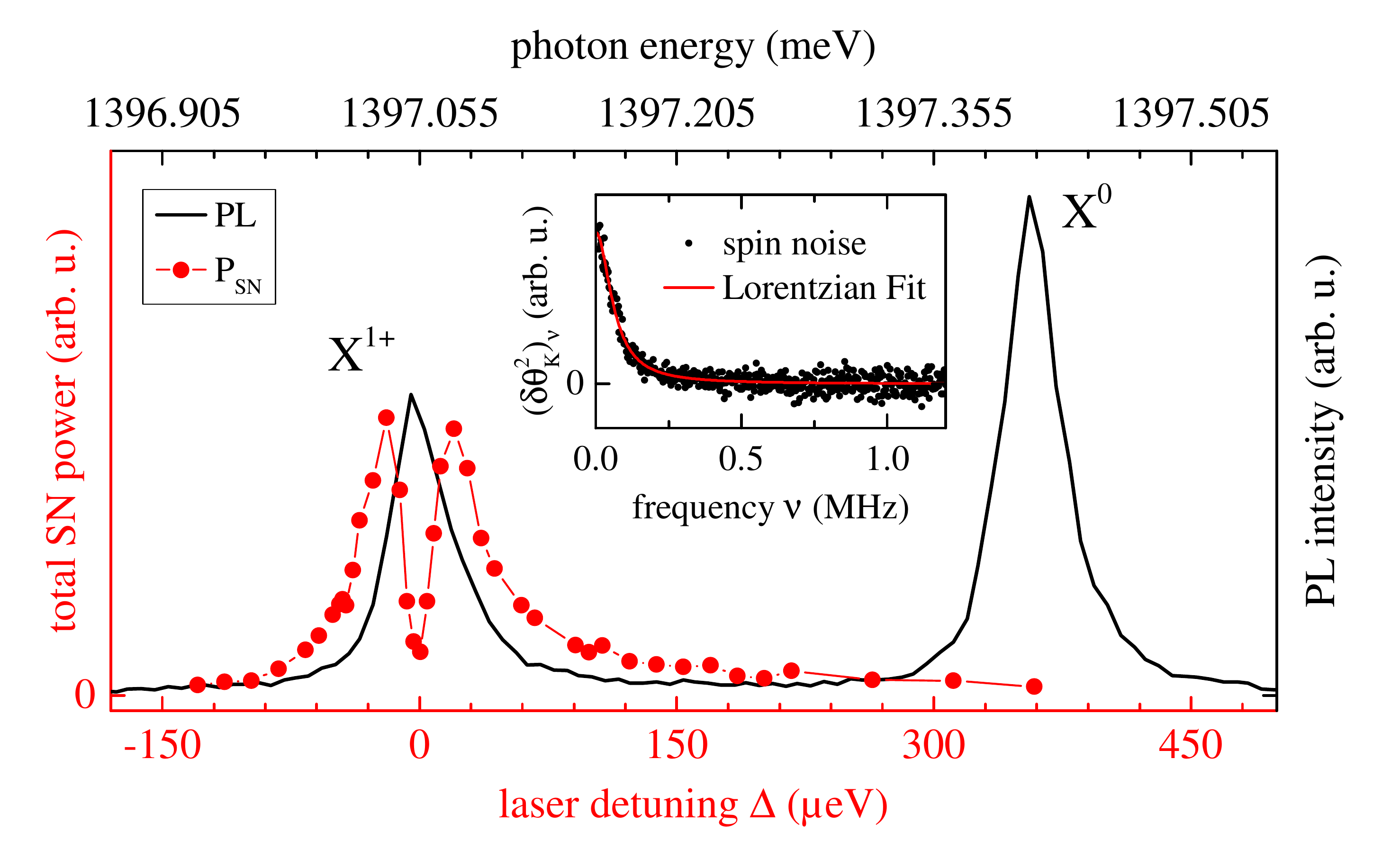}
\caption{PL spectrum of the QD showing trion ($X^{1+}$) and exciton ($X^0$) transition. The red data points depict the corresponding SN power measured as a function of laser detuning with respect to the trion resonance. The red line is a guide to the eye. The inset shows a typical SN spectrum (the black dots) at a detuning of $\Delta=-97~\mu$eV with a Lorentzian fit (the red line).}
\label{PL_SNP}
\end{figure}

The studied sample comprises a single layer of self-assembled In(Ga)As QDs grown by molecular beam epitaxy (MBE) on a (001)-oriented GaAs substrate. The QD layer has a gradient in QD density from zero to about 100 dots$/ \mu$m$^{2}$. A $p$-type doping of $\approx 10^{14}$/cm$^3$ ensures that a fraction of the QDs is occupied by a single hole~\footnote{The p-type doping is the unintentional background doping of the sample due to carbon residuals in the MBE chamber.}.
The QDs are embedded in an asymmetric GaAs $\lambda$-microcavity with 13 (top) and 30 (bottom) AlAs/GaAs Bragg mirror pairs and a $Q$-factor of about 350 determined from reflectivity measurements.
The QD microcavity is operating in the weak coupling regime, enhances the light-matter coupling, increases the ratio of SN to optical shot noise, and enables measurements in reflection geometry. The measurement setup is a low-temperature confocal microscope with two detection arms, one for photoluminescence (PL) analysis and one for SNS (see Ref.~\cite{Dahbashi2014a} and the Supplemental Material~\cite{supp}).
The QD sample is cooled down to 4.2~K and, for PL measurements, excited above the QD barrier by a cw diode laser with a photon energy of 1.59~eV.
The black solid line in Fig.~\ref{PL_SNP} shows the PL spectrum of a single QD in a sample region of very low QD density. The measured linewidth of the PL does not reflect the natural linewidth of the QD but is limited by the optical excitation power and the spectrometer resolution of about 20~$\mu$eV. All results below are measured on this specific QD but control measurements on other QDs yield consistent results.
The transition at higher photon energy is attributed to the neutral exciton ($X^0$) and the transition at lower photon energy to the positively charged trion ($X^{1+}$) (see the Supplemental Material~\cite{supp} for details),
which is in good agreement with the QD PL spectra in Ref. \cite{Schwartz2015a}. The assignment is confirmed below by spectrally resolved measurements of the SN power.

The SN of the QD is measured by a linearly polarized cw Ti:sapphire ring laser which is stabilized to a Fizeau wavelength meter and focused onto the sample to a spot with a diameter of 1~$\mu$m. The fluctuations of the Kerr rotation angle due to spin fluctuations in the single QD are measured by a polarization bridge with a low-noise balanced photo receiver. The resulting electric signal is amplified, digitized, and Fourier transformed in real time to obtain a SN power density frequency spectrum. The total SN power is obtained by integrating the SN frequency spectrum. A small longitudinal magnetic field ($B_z= 31$~mT) is applied to increase the longitudinal spin relaxation time $T_1$ and thereby improves the signal-to-noise ratio \cite{Dahbashi2014a}. The SN spectrum is isolated from the background of optical shot and electrical noise by subtracting a spectrum acquired in a purely transverse magnetic field ($B_x= 27$~mT). Here, the small transverse magnetic field efficiently suppresses the SN since the projection of the longitudinal spin component on the direction of detection is strongly reduced and the broad transverse spin component with the transversal spin relaxation time $T_2 \ll T_1$ is negligible in the measured frequency bandwidth.

The red dots in Fig.~\ref{PL_SNP} depict the integrated SN power measured as a function of probe laser detuning $\Delta$ with respect to the trion resonance for a laser intensity of $I=1.1\ \mu$W$/\mu$m$^2$. The SN power spectrum ${P_{\rm SN} (\Delta)}$ consists of two distinct maxima located symmetrically around the QD resonance with a sharp dip at $\Delta=0$ providing the evidence of the homogeneous broadening of the QD resonance \cite{Zapasskii13}. Thus, processes usually leading to inhomogeneous broadening in unbiased QDs, such as charge fluctuations in the QD environment \cite{Dahbashi2014a}, play a negligible role for this QD. The inset shows a typical SN frequency spectrum measured for a probe laser energy strongly negatively detuned from the optical resonance by $\Delta=-97~\mu$eV. This SN  spectrum has a Lorentzian line shape, and the corresponding half-width at half maximum (HWHM) yields the spin-correlation rate. In this case the QD excitation is negligible, and the HWHM is proportional to the inverse longitudinal heavy-hole spin-relaxation time $T_1^h= 1/(2 \pi \rm HWHM)$. We find from the Lorentzian fit $T_1^h = 2.51(11)~\mu$s, which evidences efficient decoupling of hole and nuclear spins in the applied magnetic field~\cite{eh_noise, Dahbashi2014a}.

\begin{figure*}
\includegraphics[width=2\columnwidth]{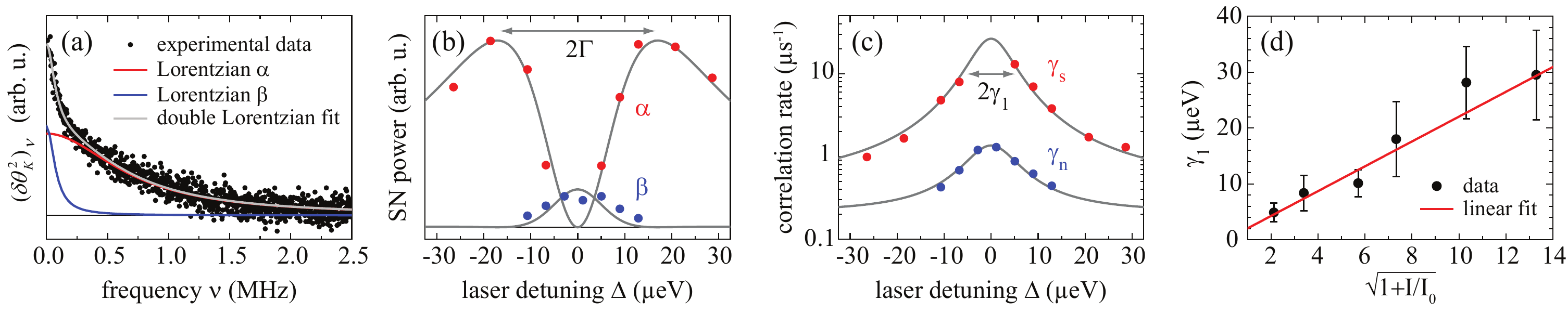}
\caption{
\label{2ndC}(a) Typical SN spectrum for a probe laser detuning of $\Delta=9~\mu$eV. The spectrum consists of two Lorentzian contributions $\alpha$ and $\beta$. (b) SN power spectrum of the $\alpha$ and $\beta$ contributions. The gray lines correspond to a global fit according to Eqs.~(\ref{SN}) and (\ref{dH}), respectively. (c) Correlation rates $\gamma_s$ ($\alpha$ contribution) and $\gamma_n$ ($\beta$ contribution) as a function of detuning. The gray lines are Lorentzian fits. (d) Intensity dependence of the trion linewidth $\gamma_1$. The red line is a fit according to Eq.~(\ref{satb}) describing saturation broadening.
}
\end{figure*}

Next, we examine the SN spectra at small detunings. The SN power has a minimum at zero detuning but does not reach exactly zero which seems, at first glance, inconsistent with the SN from a homogeneously broadened two level system. A more detailed investigation of the SN frequency spectra at a laser intensity of $1.1\ \mu$W$/\mu$m$^2$ in the quasi-resonant regime reveals: (a) an additional noise contribution and (b) the influence of strong coherent excitation of the trion. Figure~\ref{2ndC}(a) shows the SN spectrum at $9\ \mu$eV detuning as an example. In addition to the main Lorentzian denoted $\alpha$ (the red line) that dominates the SN spectrum at large laser detuning (the inset in Fig.~\ref{PL_SNP}) we observe a second Lorentzian contribution denoted $\beta$ (the blue line). The second Lorentzian has a significantly smaller width, i.e., a longer correlation time, and a significantly smaller maximal noise power compared to the $\alpha$ contribution. The respective SN power spectra of the $\alpha$ and $\beta$ contributions are depicted in Fig.~\ref{2ndC}(b).
%
%
The SN power spectrum of the $\alpha$ contribution is in excellent agreement with the expected line shape for the SN of a single hole,
\begin{align}
\label{SN}
	 P_{\rm SN,\alpha} (\Delta) = A \frac{\Delta^2}{\left(\Gamma^2+\Delta^2\right)^2},
\end{align}
where the parameter $\Gamma$ describes the width of the SN power spectrum (cf. Fig.~\ref{2ndC}(b)) and $A$ determines the amplitude of the spectrum. Note, that strong coherent excitation of the trion results in a much larger $\Gamma$ than the intrinsic linewidth $\gamma$ (see the Supplemental Material~\cite{supp} and Ref.~\cite{auger}). Indeed, a fit to the data based on Eq.~(\ref{SN}), shown by the gray line in Fig.~\ref{2ndC}(b), yields a value of $17.08(58)~\mu$eV for $\Gamma$ which is about one order of magnitude larger than the typical $\gamma$ of self-assembled QDs, in agreement with the model below.

The strong coherent driving of the system leads to the formation of new dressed states akin to trion-polaritons, representing a coherent superposition of the ground (hole) and excited (trion) states superimposed by the light~\cite{microcavities}. The eigenfrequencies of these states depend on the intensity of the probe light and the detuning. This results into a renormalization of the SN power spectrum width~$\Gamma$ (see the Supplemental Material~\cite{supp}) as shown in Fig.~\ref{2ndC}(b). Therefore the observed $\alpha$ contribution is related to the dressed-state spin noise.

The red dots in Fig.~\ref{2ndC}(c) show the detuning dependence of the $\alpha$-correlation rate which, for large detunings, is associated with the relaxation of the heavy-hole spin.
The rate strongly increases in the vicinity of the trion resonance which proves that trion excitation by the probe laser significantly affects the spin dynamics. Consistently, the correlation rates increase as well with higher laser intensities (not shown) \cite{Dahbashi2012}. The SN power and the correlation rate of the $\alpha$ contribution closest to zero detuning could not be extracted from the SN spectrum since its SN power decreases towards zero and its Lorentzian width becomes much larger than the detection bandwidth of 1.8~MHz.

\begin{figure}
\includegraphics[width=\columnwidth]{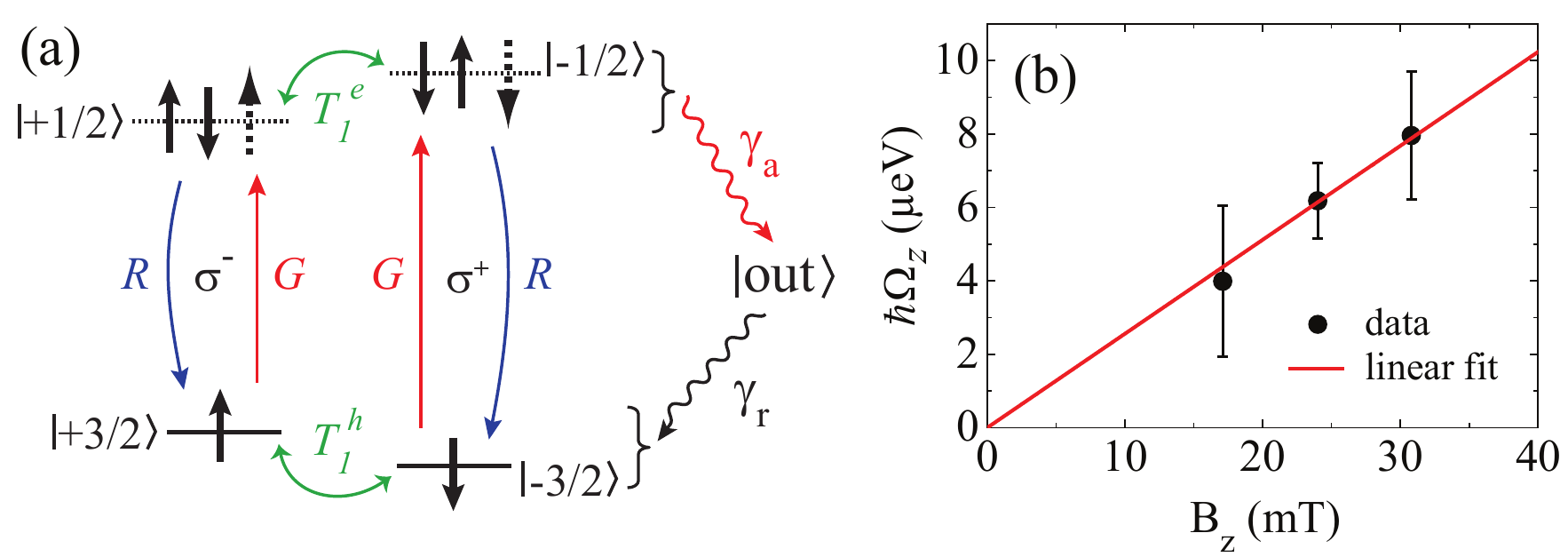}
\caption{
\label{level}(a) Sketch of the QD states and the relevant transitions. (b) Splitting of electron and hole spin states extracted from the ratio of $\alpha$ and $\beta$ SN power spectra as a function of $B_{z}$.
}
\end{figure}

The broadening of the SN power spectrum and the increase in the correlation rate of the $\alpha$ contribution can be readily explained in the framework of a four-level system shown in Fig.~\ref{level}(a).
Absorption of a $\sigma^+$ or $\sigma^-$ cavity photon by the QD results in a transition from the hole $|\pm 3/2\rangle$ spin state to the trion $|\pm 1/2\rangle$ state, respectively, as shown by the red arrows. The generation rate is given by~\cite{glazov:sns:jetp16}
\begin{equation}
  \label{eq:G}
  G=\frac{\mathcal E^2\gamma}{\gamma^2+\Delta^2},
\end{equation}
where $\mathcal E$ is the matrix element of the trion optical transition, which is proportional to the interband dipole matrix element and the electric-field amplitude in the cavity (see the Supplemental Material~\cite{supp}). The transition back to the ground state (the blue arrows in Fig.~\ref{level}(a)) can be induced either by stimulated photon emission or by trion recombination without light emission into the main cavity mode. In the weak-coupling regime the trion recombination rate $R=G+\gamma_0$ can be presented as the sum of the generation rate $G$, representing the probe-induced recombination, and the spontaneous recombination rate $\gamma_0$.
In addition, nonradiative trion recombination can result in a transition of the hole from the QD into an outer state $\rm \ket{out}$ via the Auger recombination process with a rate $\gamma_a$~\cite{auger,Klimov2000,Efros1997}. The recharging process returns a hole from $\rm\ket{out}$ back into the QD ground state with a rate $\gamma_r$. We will show below that the generation and recombination processes are by a few orders of magnitude faster than the Auger recombination, QD recharging, and spin-relaxation processes so that the steady-state occupancy of the trion state $n_{\rm tr}$ is determined by the balance of generation and recombination rates,
\begin{equation}
  n_{\rm tr}= \frac{G}{G+R}n,
\end{equation}
with $n$ being the probability to find the QD in the charged state, i.e., occupied by a hole or a trion.
The average spin-relaxation rate is the weighted sum of spin-relaxation rates of the hole in the ground state $1/T_1^h$ and the electron in the trion $1/T_1^e$~\cite{Nonresonant_nonequilibrium},
\begin{equation}
  \gamma_s=\frac{n_h}{n}\frac{1}{T_1^h}+\frac{n_{\rm tr}}{n}\frac{1}{T_1^e},
\end{equation}
where $n_h=n-n_{\rm tr}$ is the probability that the QD is in the ground state. Taking into account the dependence of the generation rate on the detuning by Eq.~\eqref{eq:G}, one can see that the dependence of $\gamma_s$ on $\Delta$ is described by a Lorentzian as shown by the gray line in Fig.~\ref{2ndC}(c). The HWHM of the Lorentzian profile $\gamma_s(\Delta)$ is determined by the trion linewidth (see the Supplemental Material~\cite{supp}) and denoted as $\gamma_1$ in the following. Figure~\ref{2ndC}(d) shows the measured $\gamma_1$ as a function of the probe laser intensity. The measured intensity dependence is perfectly described by saturation broadening (see the Supplemental Material~\cite{supp}),
\begin{equation}
  \label{satb}
  \gamma_1(I)=\gamma\sqrt{1+{I}/{I_0}},
\end{equation}
where $I/I_0=2\mathcal E^2/(\gamma\gamma_0)$. The fit of $\gamma_1$ according to Eq.~\eqref{satb} is shown as the red line in Fig.~\ref{2ndC}(d) and yields the intrinsic HWHM of the trion transition $\gamma = 2.2(23)\ \mu$eV and the saturation intensity $I_0=0.07(15)\ \mu$W/$\mu$m$^2$. The value of $\gamma$ is in good agreement with the homogeneous linewidth found in Ref. \cite{Yang2014} for positively charged QDs.

At the same time, the strong increase in the spin-relaxation rate $\gamma_s$ in the vicinity of the trion resonance is related to the fast spin relaxation of the electron in the trion, i.e., the electron spin in the trion relaxes orders of magnitudes faster than the hole spin. The fit of $\gamma_s$ for different detunings and intensities yields the longitudinal electron-in-trion spin-relaxation time $T_1^e = 32.8(32)$~ns (see the Supplemental Material~\cite{supp} for details).

Now we proceed to the detailed investigation of the $\beta$ component which is characterized by a different dependence on the detuning and a significantly lower correlation rate compared to the $\alpha$ component. The SN power spectrum of the $\beta$ contribution has a maximum at the QD resonance which suggests one of the following origins (see the Supplemental Material~\cite{supp}):
(i) spin splitting noise, which can be produced by nuclear spin fluctuations~\cite{Berski2015b}, (ii) resonance frequency fluctuations due to charge noise ~\cite{Dahbashi2014a}, and (iii) occupancy noise of the resident charge carrier in the QD.
In order to have a deeper insight into the particular mechanism we performed measurements of the SN power spectra as a function of the longitudinal magnetic field. These measurements show that the noise power of the $\beta$ component increases proportional to $B_z$ which directly excludes spin splitting noise. The trion resonance frequency fluctuations can also be excluded because they are characterized by a different detuning dependence and a shorter correlation time (see the Supplemental Material~\cite{supp}). Hence, the $\beta$ contribution is related to the occupancy noise (ON) of the QD.
At first glance, the fluctuations of the hole or trion occupancy $\delta n_h$ or $\delta n_{tr}$ alone can not produce a noise of the Kerr rotation angle $\theta_K$ because of the time-reversal symmetry~\footnote{The effects related to spatial dispersion are negligible for the system under study.}. However, the symmetry analysis (see the Supplemental Material~\cite{supp}) reveals possible contributions of the form $\theta_K \propto \delta n_h B_z$, $\delta n_{tr} B_z$. Thereby the application of a longitudinal magnetic field opens up the possibility to study the charge dynamics in the system by means of SNS.

Quantum dot occupancy noise arises when the average occupancy is smaller than unity.
The correlation rate of the QD occupancy $\gamma_n$ is determined by the rates of the transition from the trion state to an outer state and by the reoccupation of the QD by a hole, see Fig.~\ref{level}(a). The ejection of the hole out of the QD can result from the Auger process. Reoccupation can either result from hole tunneling from a nearby acceptor or capture of a free hole.
The origin of the measured occupancy noise is the finite Kerr rotation of an occupied QD as compared to the absence of any Kerr rotation for an empty QD in a longitudinal magnetic field at the trion resonance. The average Kerr rotation angle is
\begin{equation}
  \theta_K\sim \frac{\gamma_1^2-\Delta^2}{(\gamma_1^2+\Delta^2)^2}n\Omega_z,
  \label{eq:theta}
\end{equation}
where $\Omega_z=\Omega_z^e-\Omega_z^h$ is the difference between electron and hole spin splittings.
The average QD occupancy fluctuation squared is given by $\langle\delta n^2\rangle=n(1-n)$ which along with Eq.~\eqref{eq:theta} determines the dependence of the occupancy noise power on the detuning,
\begin{equation}
  \label{dH}
  P_{\rm ON,\beta} (\Delta) = A \frac{\left( \Gamma^2 - \Delta^2 \right)^2}{\left( \Gamma^2 + \Delta^2 \right)^4}\frac{1-n}{4}\Omega_z^2,
\end{equation}
cf. Eq.~\eqref{SN}.
A fit to the SN power data of the $\beta$ contribution based on this equation is shown in Fig.~\ref{2ndC}(b) as the gray line. We note however, that the exact shape of this dependence can be sensitive to the details of the charge dynamics in the outer states (see the Supplemental Material~\cite{supp}).

The comparison of the $\alpha$ and $\beta$ SN power spectra as a function of $B_z$ allows by assuming ${1-n\sim 1}$ to estimate the spin splitting $\Omega_z$, which is depicted in Fig.~\ref{level}(b). The linear dependence on $B_z$ additionally proves the correct identification of the origin of the $\beta$ contribution and yields a difference of the electron and hole $g$-factors $|g_e-g_h|\sim 4$,
which is reasonable considering the strong renormalization of the electron and hole $g$-factors in In(Ga)As QDs~\cite{Nakaoka2004,Bayer1999a}.

The blue dots in Fig.~\ref{2ndC}(c) depict the correlation rate of the reoccupation noise $\gamma_n$, which decreases strongly for increasing laser detuning. This dependence has again a Lorentzian shape and is described by
\begin{equation}
  \gamma_n=\frac{n_{\rm tr}}{n}\gamma_a+\gamma_r,
\end{equation}
where $\gamma_r$ is the reoccupation rate. Close to the optical resonance, $\gamma_n$ is dominated by the Auger recombination rate $\gamma_a$. Taking into account the saturation intensity $I_0$ and trion linewidth $\gamma$ determined from the fit of the $\alpha$ contribution we extracted the Auger rate $\gamma_a = 2.93(28)~\mu$s$^{-1}$. This is in very good agreement with the value measured in Ref.~\cite{auger} for similar QDs.
For large detuning, the correlation time is not dominated by the Auger process anymore but by the reoccupation time which is slow in our sample. Our measurements of $\gamma_n(\Delta)$ yield an estimate for the reoccupation rate of $\gamma_r = 0.207(96)~\mu$s$^{-1}$.

In conclusion, we have measured the nonequilibrium SN of a homogeneously broadened single QD inside a microcavity in the weak-coupling regime.
The presented results extend SNS to the coherent single spin dynamics investigation far beyond the fluctuation-dissipation theorem which uncovers the hidden potential to study strongly nonequilibrium but yet coherent spin dynamics of the excited states and charge dynamics in the system.
The SN of the strongly excited artificial atom shows two very distinct contributions. The dominant contribution is related to the optically driven heavy-hole trion transition which is potentially useful for spin-photon interfacing. This contribution shows the anticipated saturation broadening and a combined spin-relaxation time resulting from the ground state (hole) and the excited state (electron in the trion). The second contribution is much weaker but may be parasitic for spin-photon interfacing. This contribution results from the intrinsic loss of the heavy hole due to the small but finite nonradiative recombination rate and the subsequent reoccupation of the QD by a hole. The control of this contribution is left for future investigations and may represent an important challenge for applications (see the Supplemental Material~\cite{supp}).
We demonstrate the ability of SNS to access single charge carrier and trion spin and charge dynamics, as well as the optical properties of interband transitions under nonequilibrium conditions. We expect that this approach can become an essential powerful tool on hand for the characterization of future spin-photon interfaces and spin-based information processing devices.

\begin{acknowledgments}
We thank K. Pierz (PTB) for providing the sample and acknowledge the financial support by the joint research project Q.com-H (BMBF 16KIS00107), the German Science Foundation (DFG) (GRK 1991, OE 177/10-1), the Basis Foundation and RF President Grant No. SP-643.2015.5. Theory was developed under partial support of the Russian Science Foundation (Grant No. 14-12-501067). 
\end{acknowledgments}



%

\onecolumngrid
\vspace{\columnsep}
\begin{center}
\newpage{\large\bf {Supplemental Material to\\ ``Spin and reoccupation noise in a single quantum dot\\ beyond the fluctuation-dissipation theorem''}
}
\end{center}
\vspace{\columnsep}

This supplementary information discusses the following topics:
  \begin{enumerate}
	\item Experimental setup and PL analysis
	\item Extraction of the Auger rate and the electron spin relaxation rate
  \item Theoretical model
		\subitem 3.1 Model of spin and charge dynamics
		\subitem 3.2 Separation of timescales
		\subitem 3.3 Analysis of different probe polarization noise sources
		\subitem 3.4 Spin noise spectrum
		\subitem 3.5 Toy model of the outer states
	\item Challenges for applications due to the reoccupation noise	
  \end{enumerate}
\vspace{\columnsep}
\twocolumngrid

\renewcommand{\thepage}{S\arabic{page}}
\renewcommand{\theequation}{S\arabic{equation}}
\renewcommand{\thefigure}{S\arabic{figure}}
\renewcommand{\bibnumfmt}[1]{[S#1]}
\renewcommand{\citenumfont}[1]{S#1}

\setcounter{page}{1}
\setcounter{section}{0}
\setcounter{equation}{0}
\setcounter{figure}{0}

\renewcommand{\cite}[1]{{[}\onlinecite{#1}{]}}

\newcommand{\Tr}{\mathop{\rm Tr}\nolimits}
\newcommand{\e}{\mathrm{e}}
\renewcommand{\i}{{\rm i}}
\renewcommand{\ne}{n_{\rm e}}
\newcommand{\ntr}{n_{\rm tr}}
\newcommand{\tr}{{\rm tr}}
\newcommand{\out}{{\rm out}}
\newcommand{\aver}[1]{\left\langle #1\right\rangle}

\section{S1. Experimental setup and PL analysis}

\begin{figure}[b]
\centering
\includegraphics[width=\columnwidth]{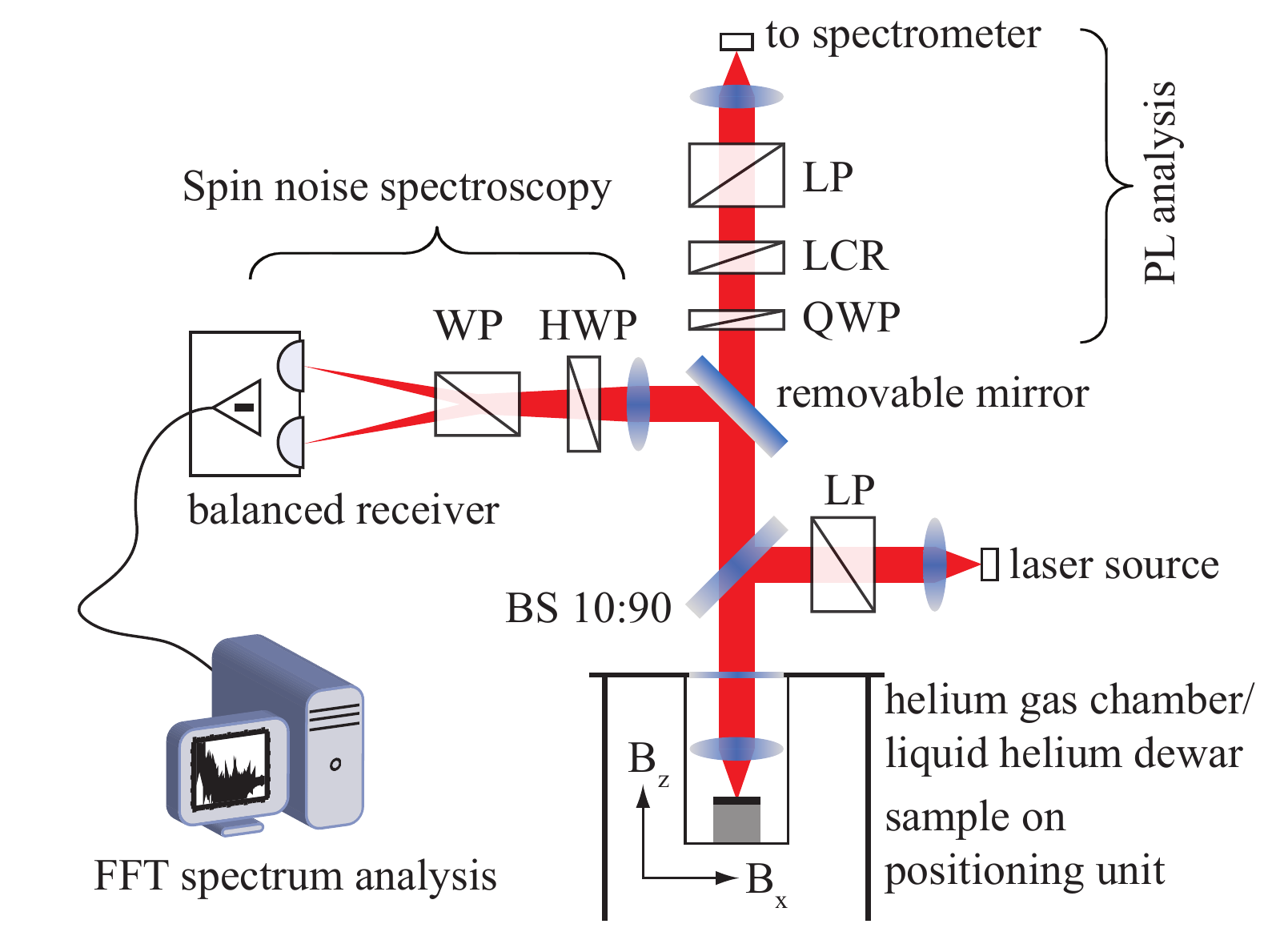}
\caption{
\label{setup} Schematic of the low-temperature confocal microscope for consecutive measurements of PL and spin noise on a single QD. A beam splitter (BS) sends 10\% of the incoming laser light down to the sample while 90\% of the light, reflected from the sample, are transmitted for detection. The quarter waveplate (QWP), liquid crystal retarder (LCR), and linear polarizer (LP) enable polarization resolved PL measurements. A removable mirror switches between PL and SN. The half wave plate (HWP) in front of the Wollaston prism (WP) enables the exact balancing of the balanced receiver for the SN measurements.
}
\end{figure}
\begin{figure}[b]
\centering
\includegraphics[width=\columnwidth]{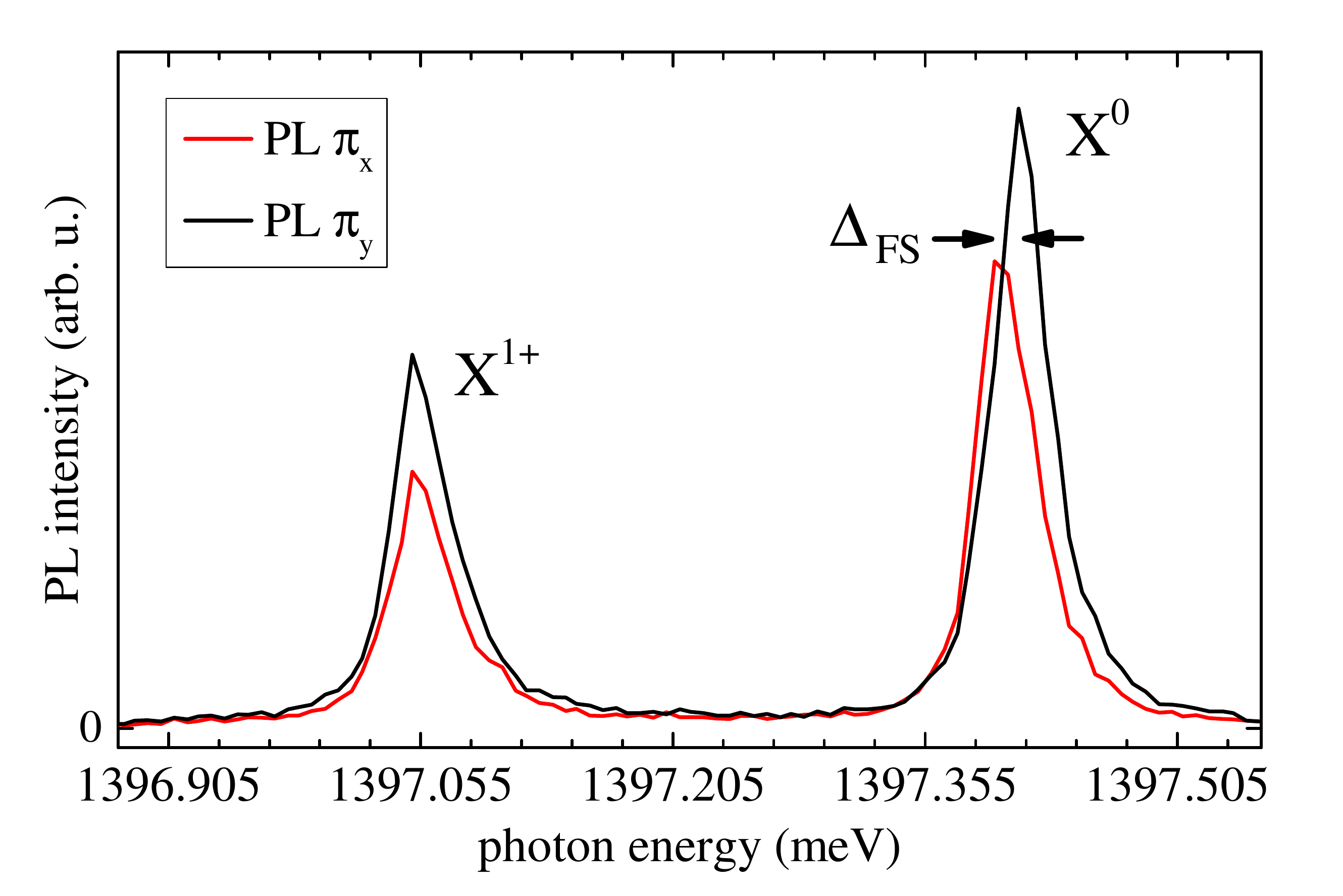}
\caption{
\label{PL} Polarization-resolved photoluminescence spectrum of the QD revealing the small fine structure splitting $\mathrm{\Delta_{FS}}$ of the exciton transition ($X^0$). The trion transition ($X^{1+}$) does not exhibit a fine structure splitting.
}
\end{figure}

Figure~\ref{setup} shows a sketch of the experimental setup. The scanning confocal microscope provides two detection
arms, one for spin noise spectroscopy (SNS) and one for photoluminescence (PL) analysis. PL measurements on a quantum dot (QD) are performed prior to SNS to determine the charge state of the QD and identify the energy of the trion transition for SN measurements.
The QD PL is analyzed with respect to the linear polarization along the two perpendicular crystal axes and spectrally resolved by a triple spectrometer with a resolution of $\mathrm{\approx 20\ \mu eV}$. Figure~\ref{PL} shows the polarization-resolved PL spectrum of the QD investigated in the main text.
The perpendicularly polarized PL components $\mathrm{\pi_x}$ and $\mathrm{\pi_y}$ reveal a small fine structure splitting, $\Delta_{\rm FS} \approx 10~\mu$eV, for the high-energy transition (at about $1397.4$~meV). The low-energy transition ($1397.06$~meV)  does not show a splitting of the polarization components.
The anisotropic exchange interaction leads for an uncharged QD to a splitting of the linearly polarized exciton eigenstates. Hence, the transition at higher photon energy is attributed to the neutral exciton ($X^0$).
The absence of the fine structure splitting is an indication for a charged QD as the exchange interaction vanishes due to the spin singlet state of the paired charge species in the trion (cf. Fig.~\ref{fig:transitions}). We exclude the occurrence of a negatively charged trion, due to the p-type doping of the sample, and attribute the transition at lower photon energy to the positively charged trion ($X^{1+}$).
Spatially resolved PL measurements, the low QD density (on the order of 1~dot/$\mu m^2$), and the statistics on other QDs ensure that both PL lines result from the same QD.
The trion resonance at 1.397~eV lies within the FWHM of the cavity resonance which is centered at $\sim 1.395$~eV at low temperatures and has a FWHM of $\sim 4$~meV.

We note that the experimental results presented in this work were obtained on a sample from the same wafer as used in Ref.~\cite{s:Dahbashi2014a}, but in a region with lower QD density which probably results in a lower density of ionized acceptors. In this region we obtained similar results for all studied QDs, in particular the inhomogeneous broadening was for all studied intensities negligible.

\section{S2. Extraction of the Auger rate and the electron spin relaxation rate}

In the main text we present the data of the correlation rates of $\alpha$- and $\beta$-contribution as a function of laser detuning (cf. Fig. 2(c)). Both correlation rates, $\gamma_s$ and $\gamma_n$, exhibit a Lorentzian detuning dependence with the HWHM corresponding to the trion linewidth $\gamma_1$. The correlation rate profiles, $\gamma_s(\Delta)$ and $\gamma_n(\Delta)$, were measured for different probe laser intensities which enables the extraction of the Auger rate and electron spin relaxation rate as their impact on $\gamma_s$ and $\gamma_n$ increases with increasing laser intensity.

To determine the Auger rate we consider the detuning-dependent part of the $\beta$ correlation rate, that is given by
\begin{equation}
\label{eq:gn}
  \gamma_n (\Delta) = \frac{G}{G+R} \gamma_a,
\end{equation}
where $G$ and $R$ are the generation and recombination rates associated with the trion transition.
Using the relations given in the main text (see discussion around Eqs.~(2)--(5)),
\begin{subequations}
  \label{eq:mt}
	\begin{equation}
	 G = (\mathcal E^2 \gamma)/(\gamma^2+\Delta^2),
	\end{equation}
	\begin{equation}
	 R = G+\gamma_0,
	\end{equation}
	\begin{equation}
	 I/I_0 = 2 \mathcal E^2/(\gamma \gamma_0),
	\end{equation}
\end{subequations}
and integrating $\gamma_n(\Delta)$, we obtain the area under the Lorentzian profile as a function of laser intensity
\begin{equation}
\label{eq:An}
  A_n(I) = \frac{\pi}{2} \frac{I \gamma}{\sqrt{I_0 (I+I_0)}}\gamma_a,
\end{equation}
where $\gamma$ and $I_0$ are the natural trion linewidth and the saturation intensity, that are determined to $\gamma = 2.2~\mu$eV and $I_0 = 0.07~\mu$W/$\mu$m$^2$ from a fit of the measured $\gamma_1(I)$ (see Fig. 2(d) in the main text).

\begin{figure}
\centering
\includegraphics[width=\columnwidth]{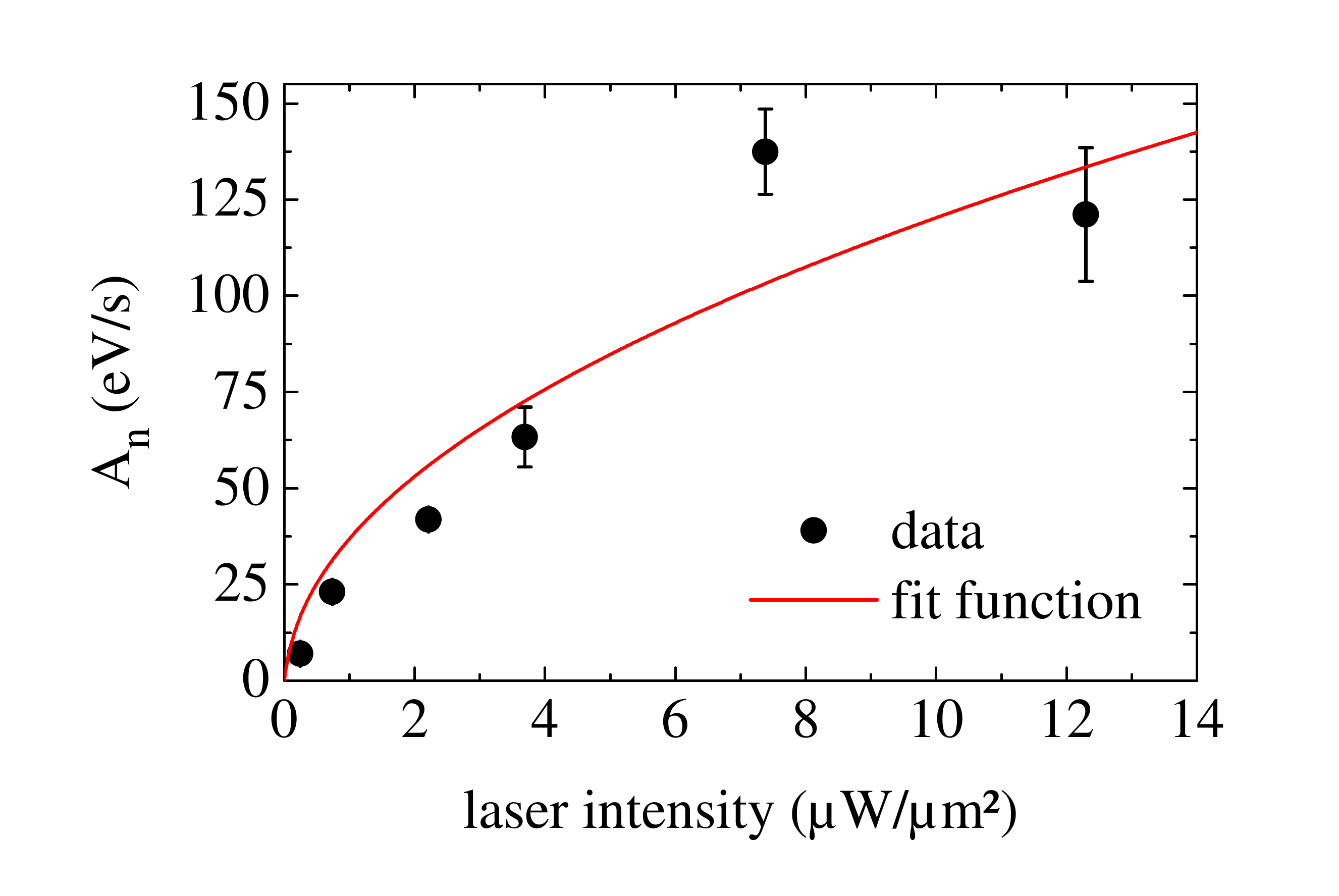}
\caption{
\label{auger}The area under the measured Lorentzian profile $\gamma_n(\Delta)$ as a function of probe laser intensity. The red line is a fit according to Eq.~\eqref{eq:An}.
}
\end{figure}
Figure~\ref{auger} shows the area of $\gamma_n(\Delta)$ obtained from a Lorentzian fit to the measured correlation rates for each laser intensity. The fit of Eq.~\eqref{eq:An} to these data yields the Auger rate $\gamma_a \approx 2.9~\mu$s$^{-1}$.

The electron spin relaxation rate is obtained in a similar way using the detuning-dependent part of the $\alpha$-correlation rate
\begin{equation}
\label{eq:gs}
  \gamma_s (\Delta) = \frac{G}{G+R} (\gamma_e-\gamma_h).
\end{equation}
Here we assume $\gamma_h\sim 0$ as the hole spin relaxation is much slower than the electron spin relaxation. The integration of $\gamma_s(\Delta)$ then yields
\begin{equation}
\label{eq:As}
  A_s(I) = \frac{\pi}{2} \frac{I \gamma}{\sqrt{I_0 (I+I_0)}}\gamma_e,
\end{equation}
where again $\gamma_e$ can be determined from a fit to the data of the measured Lorentzian profile $\gamma_s(\Delta)$ at different intensities, yielding $\gamma_e \approx 31~\mu$s$^{-1}$.

\section{S3. Theoretical Model}

\subsection{S3.1. Model of spin and charge dynamics}

\begin{figure}
  \centering
  \includegraphics[width=0.9\linewidth]{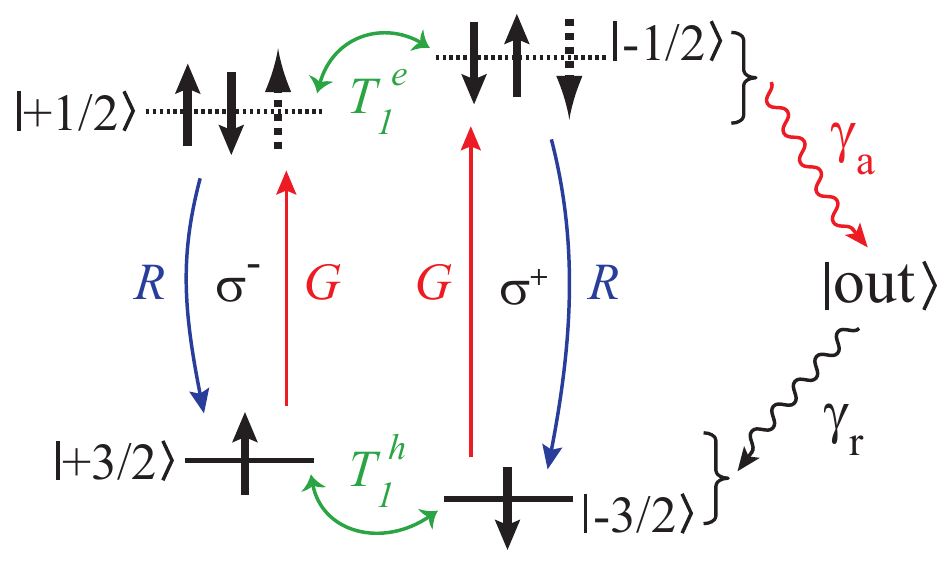}
  \caption{The sketch of the QD and outer states and transitions between them.}
  \label{fig:transitions}
\end{figure}

The theoretical description of spin dynamics in a singly charged quantum dot (QD) involves at least four spin states~\cite{s:glazov:sns:jetp16}: two heavy hole states with the spin projection $S_z^h=\pm 3/2$ on the growth axis $z$ and the two singlet trion spin states. The trion is formed of a pair of holes in the singlet spin state and an electron with the spin projection $S_z^e=\pm 1/2$. These states are shown in Fig.~\ref{fig:transitions} along with the multiple states of the resident hole outside the QD, denoted as $\left|\rm{out}\right>$.

The Hamiltonian of the system in an external magnetic field $\bm B$ in the presence of the probe light has the form
\begin{multline}
  \label{eq:H}
  \mathcal H=\hbar\omega_0 a_{s1/2}^\dag a_{s1/2} +
\frac{\hbar}{2} (\bm\Omega^h\cdot \bm \sigma)_{ss'} a_{s3/2}^\dag a_{s'3/2} \\
+ \frac{\hbar}{2} (\bm\Omega^e\cdot \bm \sigma)_{ss'} a_{s1/2}^\dag a_{s'1/2} + \left(\hbar\mathcal E_s\e^{-\i\omega t}a_{s1/2}^\dag a_{s3/2} +{\rm h.c.}\right)\:.
\end{multline}
Here the summation over the repeated indices $s,s'=\pm$ is assumed, $a_{\pm 3/2}$, $a^\dag_{\pm 3/2}$ are the anihillation and creation operators for the resident hole, $a_{\pm 1/2}$, $a^\dag_{\pm 1/2}$ are the corresponding operators of the trion, $\omega_0$ is the trion resonance frequency, $\bm\Omega^{e(h)}=\mu_B \hat g^{e(h)}\bm B/\hbar$ is the electron (hole) Larmor precession frequency with $\hat g_{e(h)}$ being the corresponding tensors of $g$-factors, $\bm\sigma$ is the vector composed of Pauli matrices, and $\mathcal E_\pm$ are the trion optical transition matrix elements in $\sigma^\pm$ polarizations, respectively, being the product of the corresponding electric field component amplitude in the QD and the transition dipole moment. We consider the weak coupling regime, so the electromagnetic field inside the microcavity can be treated classically.

 \begin{table}[b]
 \vspace{-0.3cm} \caption{Definitions of the parameters of the optical transition and the spin and charge dynamics in the QD.}
 \label{tab:params}
\begin{ruledtabular}
 \begin{tabular}{cc}
   parameter & definition \\
   \hline
   $\Delta$ & optical detuning from the trion resonance \\
   $\gamma$ & intrinsic transition dephasing rate \\
   $\gamma_0$ & trion recombination rate \\
   $\Gamma$ & renormalized $\gamma$ \\
   $\gamma_s$ & average spin relaxation rate, $\dfrac{n_h}{n}\dfrac{1}{T_1^h}+\dfrac{n_{\rm tr}}{n}\dfrac{1}{T_1^e}$ 
\\
   $\Omega$ & difference of spin splittings, $\Omega_z^e-\Omega_z^h$ \\
   $\gamma_a$ & Auger recombination rate \\
   $\gamma_r$ & QD recharging rate \\
   $\gamma_n$ & charge noise rate, $\dfrac{n_{\rm tr}}{n}\gamma_a+\gamma_r$ \\
   $\gamma_1$ & HWHM of dependencies of $\gamma_s$ and $\gamma_n$ on $\Delta$, \\
   & $\gamma\sqrt{1+2{\mathcal E}^2/(\gamma\gamma_0)}$ \\
   $\gamma_2$ & $\sqrt{\gamma_1^2+\frac{\gamma_a\gamma\mathcal E^2}{\gamma_r\gamma_0}}$ \\
 \end{tabular}
 \end{ruledtabular}
 \end{table}

The Hamiltonian accounts for the coherent processes, e.g. Rabi oscillations between hole and trion states. The incoherent processes, which are shown in Fig.~\ref{fig:transitions}, should be treated in the density matrix formalism:
\begin{equation}
\label{density:m}
  \dot\rho(t)=i[\rho(t),\mathcal H]-\mathcal L\lbrace\rho(t)\rbrace\:.
\end{equation}
Here $\rho(t)$ is the density matrix of the system, the dot denotes the time derivative,
and the Lindblad superoperator describes the spin relaxation, recombination and reoccupation processes. The charge and spin dynamics in the system are described by the set of coupled equations for the expectation values of operators $O$:
\begin{equation}
  \label{eq:aver}
  \aver{O}=\Tr\left[O\rho(t)\right].
\end{equation}
In the longitudinal magnetic field this system reads:
\begin{subequations}
  \label{eq:dynamics}
  \begin{equation}
    \label{eq:nh}
    \dot{n}_h=2\mathcal E d_x''+\gamma_0 n_\tr+\gamma_r n_\out,
  \end{equation}
  \begin{equation}
    \label{eq:ntr}
    \dot{n}_\tr=-2\mathcal E d_x''-\gamma_0 n_\tr-\gamma_a n_\tr,
  \end{equation}
  \begin{equation}
    \label{eq:Jz}
    \dot{\sigma}_h=-2\mathcal E d_y'-\frac{1}{T_1^h}\sigma_h+\gamma_0\sigma_\tr,
  \end{equation}
  \begin{equation}
    \label{eq:sz}
    \dot{\sigma}_\tr=2\mathcal E d_y'-\frac{1}{T_1^e}\sigma_\tr-\gamma_0\sigma_\tr-\gamma_a\sigma_\tr,
  \end{equation}
  \begin{equation}
    \label{eq:dxc}
    \dot{d}_x'=-\Delta d_x''-\Omega d_y'/2-\gamma d_x',
  \end{equation}
  \begin{equation}
    \label{eq:dxcc}
    \dot{d}_x''=\mathcal E(n_\tr-n_h)/2+\Delta d_x'-\Omega d_y''/2-\gamma d_x'',
  \end{equation}
  \begin{equation}
    \label{eq:dyc}
    \dot{d}_y'=\mathcal E(\sigma_h-\sigma_\tr)/2-\Delta d_y''+\Omega d_x'/2-\gamma d_y',
  \end{equation}
  \begin{equation}
    \label{eq:dycc}
    \dot{d}_y''=\Delta d_y'+\Omega d_x''/2-\gamma d_y''.
  \end{equation}
\end{subequations}
Here $n_{h}=\aver{a_{+3/2}^\dag a_{+3/2}+a_{-3/2}^\dag a_{-3/2}}$ and $n_{\tr}=\aver{a_{+1/2}^\dag a_{+1/2}+a_{-1/2}^\dag a_{-1/2}}$
are the populations of the QD ground state and the trion state, respectively. Similarly
$\sigma_{h}=\aver{a_{+3/2}^\dag a_{+3/2}-a_{-3/2}^\dag a_{-3/2}}$ and $\sigma_{\tr}=\aver{a_{+1/2}^\dag a_{+1/2}-a_{-1/2}^\dag a_{-1/2}}$
are the $z$ spin polarizations of the corresponding states. $\Omega=\Omega_z^e-\Omega_z^h$ is the total spin splitting of the optical transitions in $\sigma^+$ and $\sigma^-$ polarizations. The quantities proportional to the components dipole moment are defined in the canonical basis~\cite{s:Varshalovich} as
\begin{equation}
  d_x=\frac{-d_++d_-}{\sqrt{2}},
  \quad
  d_y=-\i\frac{d_++d_-}{\sqrt{2}},
\end{equation}
where $d_\pm=\aver{a_{\pm3/2}^\dag a_{\pm 1/2}}$ and one or two primes in Eqs.~\eqref{eq:dynamics} denote the real or imaginary parts of the corresponding expectation values, respectively. Finally we consider linearly polarized probe light, so that $\mathcal E_\pm=\mp\mathcal E/\sqrt{2}$ with $\mathcal E$ being a real constant. Therefore the dipole moment components denoted by one prime are in phase with the probe beam, and two primes stands for the component shifted by $90^\circ$. The Faraday and ellipticity signals are proportional, respectively to $d_y''$ and $d_y'$~\cite{s:yugova09}.

In Eqs.~\eqref{eq:dynamics} we introduce the hole and trion spin relaxation rates $T_1^h$ and $T_1^e$, respectively,
and assume, that the nuclear spins are effectively decoupled from the hole and electron spins due to the strong magnetic field~\cite{s:eh_noise}, so the hyperfine interaction can be included in the phenomenological hole and trion spin relaxation rates~\cite{s:Glazov2015}. For the reader's convenience we summarize the definitions of the various dephasing rates in Tab.~\ref{tab:params}. The trion recombination rate, unrelated to the cavity photon emission, is denoted as $\gamma_0$. The Auger process leads to the recombination of the trion accompanied by the escape of the hole from the QD with the rate $\gamma_a$. Finally $\gamma_r$ stands for the recharging of the QD, i.e. it is the rate of transitions from the outer states to the hole ground state in the QD. The spin relaxation in the outer states is assumed to be very fast, so that the hole returns to the QD spin unpolarized. The direct transitions from the QD ground state to the outer states are neglected because of the strong spatial confinement. In addition the capture of an electron by the QD can in principle lead to the emptying of the QD. We neglect this process since the QD sample is $p$-type and the optical excitation is quasi-resonant to the QD transition.
The hole-trion dephasing rate is denoted as $\gamma$. It is in general case greater or equal to $(\gamma_0+\gamma_a)/2+1/T_2^h+1/T_2^e$ with $T_2^{h,e}$ being the hole and trion transverse spin relaxation rates.

The system of Eqs.~\eqref{eq:dynamics} is not complete, because one has to consider also the charge dynamics in the outer states. The parameter $n_{out}$ in Eq.~\eqref{eq:nh} describes the occupancy of some outer states with the average reoccupation rate $\gamma_r$. In principle there could be multiple outer states with different transition rates to the QD ground state. Moreover, transitions between the outer states can take place as well.
In this paper we do not study the charge dynamics outside the quantum dot. Accounting for them can in principle lead to a renormalization of the parameters $\gamma_r$ and $\gamma_a$, as well as to a slight modification of the SN power spectrum. The simplest model, assuming a single outer state, is discussed in Sec.~\hyperref[sec:discussion]{\textcolor{blue}{S3.5}}.

\subsection{S3.2. Separation of timescales}

In the typical QD systems, and in particular in the sample under study, the trion generation and recombination are much faster, than all the other timescales. This means that during the time $\sim\gamma_0^{-1}$ the quasi-equilibrium is established with respect to the generation and recombination processes. This quasi steady state parametrically depends on the total QD population $n=n_h+n_\tr$ and the total spin $S_z=(\sigma_h+\sigma_\tr)/2$. These quantities are preserved during the trion generation and recombination, and their dynamics are described by slow spin relaxation and Auger processes.

Hence, from Eq.~\eqref{eq:dynamics} under the experimentally relevant assumptions
\begin{equation}
    \gamma_0,\Delta\gg\Omega,\gamma_e,\gamma_h,\gamma_a,\gamma_r
\end{equation}
one finds that in the quasi steady state (c.f. Eq.~(3))
  \begin{equation}
    \label{eq:Sze}
    \frac{n_\tr}{n}=\frac{\sigma_\tr}{2S_z}=\frac{G}{G+R},
  \end{equation}
where we have introduced the generation rate $G$ (Eq.~(2)) and the recombination rate $R=\gamma_0+G.$
Note that in the limit $\mathcal E\to0$ the generation rate vanishes and $R=\gamma_0$.

\subsection{S3.3. Analysis of different probe polarization noise sources}

The Faraday rotation is determined by the component of the dipole moment $d_y''$. In the quasi steady state it can be presented as
  \begin{equation}
    \label{eq:dy2}
    d_y''=\frac{{\mathcal E}\Delta}{\gamma_1^2+\Delta^2}S_z+\frac{{\mathcal E}(\gamma_1^2-\Delta^2)}{4(\gamma_1^2+\Delta^2)^2}n\Omega,
  \end{equation}
where $\gamma_1=\gamma\sqrt{1+2{\mathcal E}^2/(\gamma\gamma_0)}$ (cf Eq.~(5) in the main text) is the trion linewidth renormalized by the saturation broadening effect. From this expression one can see, that the Faraday rotation can be induced by (i) hole or trion spin polarization and (ii) longitudinal magnetic field. We note that the second spin-independent term can contain only the odd powers of $\Omega$. This follows from the time-reversal symmetry (Onsager principle): The Faraday rotation angle is proportional to the antisymmetric (with respect to permutations of the cartesian subscripts) part of the dielectric constant tensor which can be proportional to the odd-at-time-reversal physical quantities such as magnetic field, spin, or electric current~\cite{s:ll8_eng,s:PhysRevB.95.045406}. The effects related to the spatial dispersion, i.e., to the light wave vector, can be neglected for this system.

Experimentally the spin noise is studied in reflection geometry. The rotation of the polarization plane of the reflected light (Kerr rotation) can be in principle contributed by the dipole moment components $d_y'$ and $d_y''$ as well, depending on the position of the QD inside the cavity and the thickness of the cap layer~\cite{s:yugova09}. Indeed we have observed somewhat different SN power spectra for different QDs. In this paper we focus mainly on a particular QD, where the Kerr rotation profile is consistent with the analysis of $d_y''$ as for the Faraday rotation.

Analysis of Eq.~\eqref{eq:dy2} reveals multiple possible sources of Kerr rotation fluctuations and yields the corresponding SN power spectra.

The first source is the spin noise. The corresponding SN power spectrum is
\begin{equation}
  \label{eq:contrib_SN}
  (\delta\theta_K^2)_{SN}\sim\left(\frac{2{\mathcal E}\Delta}{\gamma_1^2+\Delta^2}\right)^2\aver{\delta S_z^2}_0,
\end{equation}
cf. Eq.~(1) of the main text. The angular brackets with the subscript 0 denote the quantum mechanical average with the steady state density matrix, or equivalently the temporal average. Eq.~\eqref{eq:contrib_SN} describes the usual spin-induced Kerr rotation noise profile~\cite{s:Zapasskii13}, characterized by the dip exactly at the trion resonance, $\Delta=0$. This contribution describes the $\alpha$-contribution in the observed spectra. We note that
\begin{equation}
  \label{eq:Sz20}
  \aver{\delta S_z^2}_0=\overline{n}/4,
\end{equation}
where $\overline{n}\equiv\aver{n}_0$ is the average steady state QD occupancy, which in principle also depends on the detuning and the probe power. This dependence is determined by the Auger process, as well as by the transitions between outer states and the reoccupation process. Detailed description of this dynamics is beyond the scope of the present study, and since this dependence does not qualitatively change the shape of the SN power spectrum, we phenomenologically describe the spectrum by Eq.~(1) of the main text, where the parameter $\Gamma$ can be different from $\gamma_1$. The simplest model describing the renormalization of $\gamma_1$ is presented in Sec.~\hyperref[sec:discussion]{\textcolor{blue}{S3.5}}.

The QD occupation noise also gives rise to the Kerr rotation fluctuations as
\begin{equation}
  \label{eq:contrib_ON}
  (\delta\theta_K^2)_{ON}\sim\frac{{\mathcal E}^2(\gamma_1^2-\Delta^2)^2}{4(\gamma_1^2+\Delta^2)^4}\Omega^2\aver{\delta n^2}_0.
\end{equation}
This contribution appears only in the presence of the longitudinal magnetic field. Moreover, it requires that there is a nonzero average of QD occupancy fluctuation $\aver{\delta n^2}_0$, i.e., a finite probability to find the QD in the empty state. This contribution is maximum at the trion resonance and the dependence on the probe frequency is similar to a Gaussian profile. The QD occupancy fluctuations are responsible for the $\beta$-contribution in the observed spectra. The dependence of $\aver{\delta n^2}_0=\overline{n}(1-\overline{n})$ on the detuning and renormalization of $\gamma_1$ in this contribution are also discussed in Sec.~\hyperref[sec:discussion]{\textcolor{blue}{S3.5}}.

For the sake of completeness we present the other possible contributions to the Kerr rotation noise. The nuclear spin fluctuations give rise to the additional splitting $\delta\Omega_n$ of the trion transitions in $\sigma^+$ and $\sigma^-$ polarizations~\cite{s:Berski2015b}. The nuclear induced Kerr rotation noise power is given by
\begin{equation}
  (\delta\theta_K^2)_{NSN}\sim\frac{{\mathcal E}^2(\gamma_1^2-\Delta^2)^2}{4(\gamma_1^2+\Delta^2)^4}\overline{n}\aver{\delta \Omega_n^2}_0.
\end{equation}
This contribution is characterized by the same dependence on the detuning as $(\delta\theta_K^2)_{ON}$, but (i) does not depend on the magnitude of the external magnetic field in contrast to the results presented in the main text and (ii) has typically a  smaller amplitude. Indeed the typical spin splitting induced by the Overhauser field is $\sim 0.5~\mu$eV, while in the applied magnetic field $B_z=31$~mT the spin splitting is expected to be a bit larger. Therefore we conclude that this contribution is not observed in our experiment.

The charge noise in the environment of the QD leads to the fluctuations of the resonance frequency $\omega_0$, and as a result to fluctuations of the detuning $\Delta$. This in turn leads to the fluctuations of both terms in Eq.~\eqref{eq:dy2}. The corresponding contribution to the SN power spectrum reads
\begin{multline}
  \label{eq:contrib_CN}
  (\delta\theta_K^2)_{CN}\sim
  \left[
    \frac{(2\mathcal E)^2(\gamma_1^2-\Delta^2)^2}{(\gamma_1^2+\Delta^2)^4}\aver{S_z^2}_0
  \right.\\    \left.
    +
    \frac{(\mathcal E\Delta)^2(3\gamma_1^2-\Delta^2)^2}{(\gamma_1^2+\Delta^2)^6}\Omega^2\overline{n}
  \right]
  \aver{\delta\omega_0^2}_0.
\end{multline}
The first line of this expression gives rise to the same SN power spectrum, as $(\delta\theta_K^2)_{ON}$ and $(\delta\theta_K^2)_{NSN}$, but it requires the presence of spin fluctuations, $\aver{S_z^2}_0$. Therefore the corresponding correlation time can not be longer than the spin relaxation time. Since experimentally the correlation time of the $\beta$-component is longer than that of the $\alpha$-component, we conclude that this contribution was not observed. Finally, the second line of Eq.~\eqref{eq:contrib_CN} describes the contribution to Kerr rotation noise, which has a dip at zero detuning. Hence, it is inconsistent with the observed $\beta$-contribution SN power spectrum. The fact that the charge noise in the surrounding of the quantum dot does not provide additional contribution to the SN power spectra is also supported by the observation of the SN power spectrum $(\delta\theta_K^2)_{SN}$, with the dip at the QD resonance, in contrast to Ref.~\cite{s:Dahbashi2014a}.

\subsection{S3.4. Spin noise spectrum}
\label{sec:SNS}

In the previous subsection we have shown, that $\alpha$ and $\beta$-contributions to the Kerr rotation noise are related to the spin and QD occupation fluctuations, respectively. In order to describe the corresponding SN frequency spectra we consider slow dynamics of $S_z$ and $n$ assuming that the quasi-equilibrium is established with respect to the generation and recombination processes in the QD, as described in the main text.

Taking into account Eq.~\eqref{eq:Sze} one finds from Eqs.~\eqref{eq:Jz} and~\eqref{eq:sz}
\begin{equation}
  \dot{S_z}=-\gamma_sS_z,
\end{equation}
where $\gamma_s$ is given by Eq.~(4) of the main text. In order to find the SN frequency spectrum we apply the Langevin
method~\cite{s:ll5_eng,s:NoiseGlazov}. The spin fluctuation obeys the equation
\begin{equation}
  \dot{\delta S}_z(t)+\gamma_s \delta S_z(t)=\xi_s(t),
\end{equation}
where the Langevin force $\xi_s(t)$ is described by the correlator
\begin{equation}
  \aver{\xi_s(t)\xi_s(t')}_0=2\gamma_s\aver{\delta S_z^2}_0\delta(t-t').
\end{equation}
This equation along with Eq.~\eqref{eq:Sz20} yields the spin noise spectrum
\begin{equation}
  \aver{\delta S_z^2}=\frac{\overline{n}}{2}\frac{\gamma_s}{\gamma_s^2+\omega^2}.
\end{equation}
Hence, the SN frequency spectrum, as usual, has the form of a Lorentzian, centered at zero frequency and has a HWHM equal to $\gamma_s$.

Similarly the slow dynamics of QD population are described by
\begin{equation}
  \label{eq:nout}
  \dot{n}=-\frac{G\gamma_a}{R+G}n+\gamma_r n_{out},
\end{equation}
where we have used Eqs.~\eqref{eq:nh}, \eqref{eq:ntr} and~\eqref{eq:Sze}. We recall, that $n_{out}$ is the population of the states, which predominantly contribute to the QD reoccupation. Therefore in order to describe the charge dynamics in the QD one has to adopt some model of the outer states.

\subsection{S3.5. Toy model of the outer states} \label{sec:discussion}

The detailed investigation of charge dynamics outside the QD is beyond the scope of this paper. In this section we present the simplest model, in which all the outer states are reduced to one quantum level, and demonstrate the qualitative consequences of QD emptying and reoccupation dynamics.

Provided all the outer states are reduced to a single state, its population satisfies the equation
\begin{equation}
  \dot{n}_{out}=\gamma_an_\tr-\gamma_r n_{out},
\end{equation}
and the particle conservation states that $n_{out}=1-n$. Taking into account Eq.~\eqref{eq:nout} one finds the steady-state QD occupancy
\begin{equation}
  \label{eq:n_steady}
  \overline{n}=\frac{\gamma_r(G+R)}{\gamma_r(G+R)+\gamma_a G}.
\end{equation}
When Auger recombination is slow, $\gamma_a\ll\gamma_r$, the population is unity independent on the generation and recombination rates. In realistic systems the opposite limit is realized, $\gamma_a\gg\gamma_r$~\cite{s:auger}, in which even a small excitation $G\sim R(\gamma_r/\gamma_a)$ leads to the efficient emptying of the QD.

Substituting Eq.~\eqref{eq:n_steady} into Eqs.~\eqref{eq:contrib_SN} and~\eqref{eq:contrib_ON} one obtains the following SN power spectra
\begin{equation}
  \label{eq:theta1}
  (\delta\theta_K^2)_{SN}\sim\frac{2\left({\mathcal E}\Delta\right)^2}{(\gamma_1^2+\Delta^2)(\gamma_2^2+\Delta^2)},
\end{equation}
\begin{equation}
  \label{eq:theta2}
  (\delta\theta_K^2)_{ON}\sim\frac{({\mathcal E}^2\Omega)^2(\gamma_2^2-\gamma_1^2)(\gamma_1^2-\Delta^2)^2}{4(\gamma_1^2+\Delta^2)^3(\gamma_2^2+\Delta^2)^2},
\end{equation}
where we have introduced
\begin{equation}
  \gamma_2=\sqrt{\gamma_1^2+\frac{\gamma_a\gamma\mathcal E^2}{\gamma_r\gamma_0}}.
\end{equation}
In general case one can see that $\gamma_2>\gamma_1$. Therefore Eqs.~\eqref{eq:theta1} and~\eqref{eq:theta2} describe the SN power spectra with the characteristic width larger than $\gamma_1$ because of the fast Auger recombination. The physical reason for the broadening is the fast emptying of the QD for small generation rates, which leads to the strong suppression of spin noise at small detunings and the enhancement of the relative contribution at large detunings.

We find that the presented toy model does not quantitatively describe the observed SN power spectra. Therefore we describe the spin noise power spectra by the expressions Eqs.~(1) and~(7) of the main text, which can be obtained from Eqs.~\eqref{eq:contrib_SN} and~\eqref{eq:contrib_ON} by the phenomenological renormalization of $\gamma_1$ to $\Gamma$.

Note that, despite the charge dynamics outside the QD significantly modify the SN power spectra, they do not affect the width of the SN frequency spectra, Eqs.~(4) and~(8). Indeed the spin relaxation rate is obtained in Sec.~\hyperref[sec:discussion]{\textcolor{blue}{S3.4}} without any assumptions about outer states and the occupancy fluctuations are described by
\begin{equation}
  \dot{\delta n}(t)+\gamma_n\delta n(t)=\xi_n(t),
\end{equation}
where $\gamma_n$ is given by Eq.~(8). Although the second term in Eq.~(8) depends on the model of dynamics in outer states, the main contribution, given by the first term is not dependent on these dynamics. As one can see in Fig.~2(c), the Auger related contribution to the occupancy correlation time dominates at small detunings.
We recall that in the Langevin method the random force $\xi_n(t)$ satisfies
\begin{equation}
  \aver{\xi_n(t)\xi_n(t')}_0=2\gamma_n\aver{\delta n^2}_0\delta(t-t'),
\end{equation}
and is introduced formally in order to support the average of squared fluctuation $\aver{\delta n^2}_0$.

Therefore the increase of the rates $\gamma_s$ and $\gamma_n$ is described by the Lorentzian profile with the HWHM given by $\gamma_1$ independent on the dynamics in outer states. This fact allows for the reliable measurement of hole and trion spin relaxation times, as well as the Auger rate, demonstrated in this work.

\section{S4. Challenges for applications due to the reoccupation noise}

The Auger effect leading to the observed reoccupation noise is intrinsic and therefore present in any quantum dot based spin-photonic structure utilizing trions. The Auger rate of the trion recombination is by an order of magnitude faster than the spin dephasing rate of the undisturbed hole and inevitably leads to the escape of the resident charge carrier from the QD. The experiment presented in the manuscript shows that the reoccupation of the QD after the Auger process is very slow in unbiased, high quality samples which leads to dead-times that strongly affect the fidelity of spin-photon interfaces, devices aimed at spin-photon entanglement, and spin-photon quantum information devices in general. The Auger effect is also present in charge-tunable quantum dot structures as shown in Ref. \cite{s:auger}, i.e., even these structures, which offer certain benefits compared to unbiased QDs, can not eliminate the Auger effect.

In theory, the  Auger effect could be reduced by reducing the density of hole states in the valence band at an energy corresponding to the trion transition energy. However, the band offset between active and barrier material is not high enough in technologically relevant semiconductors. In fact, the trion Auger recombination is additionally related to phonon emission and can hardly be avoided in any semiconductor structure.

Charge-tunable structures can be used to recharge the QD after a harmful Auger recombination immediately by altering the bias-dependent tunneling time. However, immediate recharging is linked to a strong interaction of the localized hole with the Fermi sea in the back gate which degrades the spin coherence time \cite{s:cotunneling}. Therefore, also charge-tunable structures will not eliminate the problem but have to be balanced between fast recharging and long hole spin coherence times.
In any case, the intrinsic non-radiative recombination and the resulting necessity of recharging pose an important challenge for the fabrication of efficient spin-photon interfaces.


%

\end{document}